# Interacion of Heavy Metal Ions with C-Phycocyanin: Binding Isotherms and Cooperative Effects

Eteri Gelagutashvili

E.L. Andronikashvili Institute of Physics, 6
Tamarashvili str., Tbilisi, 0177, Georgia
E.mail: gel@iphac.ge
gelaguta@yahoo.com

Abstract

The binding constant of copper(II) ions to C-PC were determined at different ionic strengths from binding isotherms by equilibrium dialysis and flame atomic absorption spectroscopy. The binding constants of 2, 20, 50 mM ionic strengths were found to be $1.1 \times 10^5$ $M^{-1}$, $0.81 \times 10^5$ $M^{-1}$, $0.74 \times 10^5$ $M^{-1}$ respectively.

Fluorescence and absorbtion spectroscopy provides insight of metal-C-phycocyanin interactions. Fluorescence measurements demonstrate C-PC quenching of heavy metal ions emission intensities. Stern-Volmer quenching constants were obtained from the linear quenching plots. Blue shifts in the fluorescence spectra were observed during metal binding to C-PC. It was shown, that between bound metal ions in C-PC there exists positive cooperativity. Plots constructed from fluorescence spectral data gave binding constants: $11.4 \times 10^5 M^{-1}$(mercury(II)); $5.2 \times 10^5 M^{-1}$(silver(I)); $2.8 \times 10^5 M^1$(lead(II)); $1.0 \times 10^5 M^{-1}$(Chromium(III)); $0.6 \times 10^5 M^{-1}$ (copper(II));

## Introduction

Heavy metal ions are a major cause of environmental pollution. Many of these are demanded in trace quantities as nutrients and only become toxic at higher concentration.
Quality control for Spirulina as a food includes microbiological standard tests, chemical composition test and test for heavy and toxic metals. *Spirulina platensis* is a filamentous cyanobacterium that is biotechnologically important due its high nutritional value[1]. The nutritional value derives from its high protein content (about 70%) and its type of lipids (g-linolenic acid). Although this ancient algae has been eaten for centuries by traditional people, it was only rediscovered by scientists 30 years ago. In fact, the United Nations and the World Health Organization recommend spirulina as safe and nutritious for children. It is known, that *S.platensis* has higher sorption capacity for cadmium[2], but interaction of toxic and heavy metal ions with basic protein C-phycocyanin (C-PC) is not defined.
Potential role of C-PC in hepatoprotection, anti-inflammatory and anti-arthritic action postulated in [3]. C-PC exerts photodynamic action on tumor cells and might be used as one of the promising candidates for the reagent of new type in photodynamic therapy[4]. C-PC from *S.platensis* inhibited growth of human leukemia K562 cells[5]. The virtual lack of toxicity of phycocyanin suggests that it can be used in the treatment of the Alzheimer's and Parkinson's diseases. Phycocyanin reduced experimental status epilepticus, suggesting possible therapeutic intervention in the treatment of some forms of epilepsy[6]. It was suggested that phycocyanin might be responsible for the suppression of renal toxicity induced by inorganic mercury and *cis*-platin[7].

In this work the interaction of metal ions with C-PC was investigated using optical and thermodynamic methods.

## Materials and methods

C-PC was isolated from laboratory cultures of the blue-green alga Spirulina platensis, according to Teale and Dale[8]. The purity of the protein was assessed from the ratio of the absorbances at $\lambda=615nm$ and $\lambda=280nm$ ($A_{615}/A_{280}>4$). The concentration of C-PC was determined by UV/Vissible spectroscopy using a value of $\varepsilon_{\lambda=615nm}=279\,000$ $M^{-1}cm^{-1}$ for the absorption coefficient. Other reagents (metal ions), all of analytical grade were prepared in double distilled water. Fluorescence spectra were measured in $1cm^3$ quartz cells using fluorescence spectroscopy. The solution were excited at 488nm
and the fluorescence was monitored at 635nm.
Equilibrium dialysis experiments were performed in a two-chambered Plexiglass apparatus. The chamber's capacity was 5 ml. The membrane's thickness was 30 μm (Visking). The initial metal concentration was varied within the range $10^{-6}$-$10^{-4}$M. Samples were analyzed by flame atomic-absorption spectrophotometry. All experiments were carried out at $23^0C$ at different ionic strengths.

## Results and Discussions

The fluorescence spectra of C-PC in the presence of $Hg^{2+}$ ions are shown as an example in Fig.1. The intensity of C-PC fluorescence, observed in the range 580-690 nm, was measured as a function of the added metal concentration. Metal free form of C-PC has a fluorescence maximum at 635 nm. The increase of the metal concentration causes the decrease in the peak amplitude and was accompanied by a blue shift of fluorescence peak (5-15nm). Fig.2, where normalized fluorescence spectra of C-PC in the absence and in the presence (2) of Hg ion are shown, also confirms this.

It is clear from Fig. 1, that quenching has not been saturated even at high concentrations of metal ions. The absence of saturation in the quenching plot even when the binding is saturated indicates that only a fraction of the binding sites quenches the fluorescence.

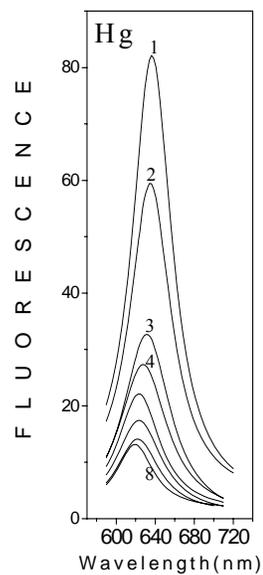

**Fig.1.** Fluorescence spectra of C-PC (0,4 µM) in the presence of different concentrations of Hg(II) ions. 1→8 [Hg] = 0 →7 (µM)

.

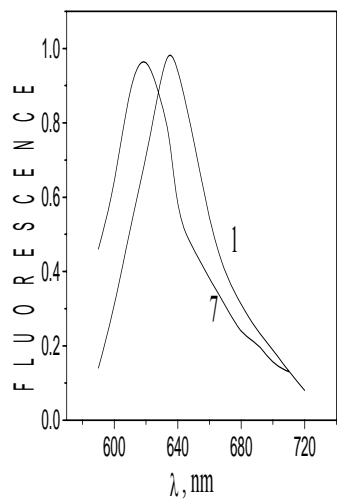

Fig.2. Normalized fluorescence spectra of C-PC in the absence (1) and in the presence (7) of Hg ion.

Relative fluorescence intensities of C-PC in the presence of various metal cations are shown in Fig. 3.

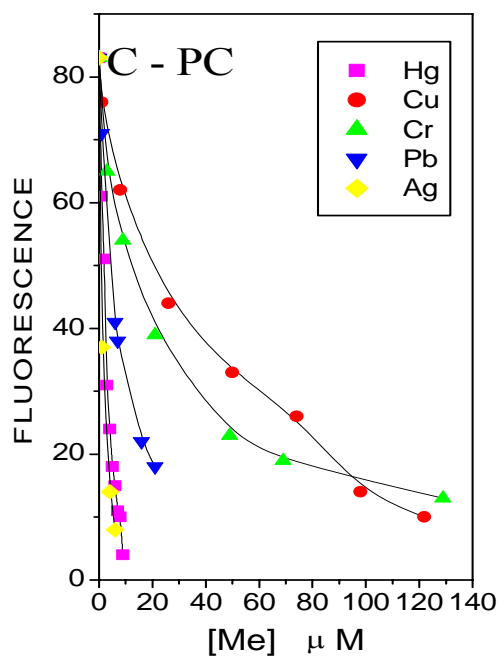

Fig.3. Relative fluorescence intensities of C-PC (0.4µM) in the presence of various metal cations .

The concentrations of metal ions that quench 50% of C-PC fluorescence intensity (IC50) are given in Table 1.

**Table 1.** Quenching of C-PC fluorescence by various metal Ions.

| Metal ion | Hg(II) | Ag(I) | Pb(II) | Cu(II) | Cr(III) |
|---|---|---|---|---|---|
| IC50(µM) | 2.5 | 3 | 6 | 26 | 21 |

Hg(II) ions quench the fluorescence more strongly than Ag(I) and Pb(II) and the lowest quenching efficiency were observed for Cu(II) and Cr(III) ions, consistent with low affinity of C-PC for these ions.

The quantitative analysis of quenching efficiency is based on the classic Stern-Volmer equation $(I_0/I = 1 + K_{SV}[C])$, where $I_0$ and $I$ are the fluorescence intensities without and with metal ions, $[C]$ is the metal cation concentration and $K_{SV}$ is the collision quenching constant (Stern-Volmer), that is determined by the excited state lifetime and by the velocity constant of quenching collisions.

$K_{SV}$ values for C-PC fluorescence quenching by various metal ions are presented in Table 2.

Table 2. $K_{SV}$ values for C-PC fluorescence quenching by various metal ions.

| Metal ion | Hg(II) | Ag(I) | Pb(II) | Cu(II) | Cr(III) |
|---|---|---|---|---|---|
| $K_{SV} \times 10^5$ M$^{-1}$ | 10 | 3 | 2 | 0.3 | 0.5 |

The fluorescence titration data on C-PC complexes with toxic and heavy metal ions (Hg(II), Cu(II), Cr(III), Pb(II), Ag(I)) were plotted as binding isotherms and then analyzed by the Scatchard and Hill graphical methods ( Fig.4  Hg(II)-C-PC ).

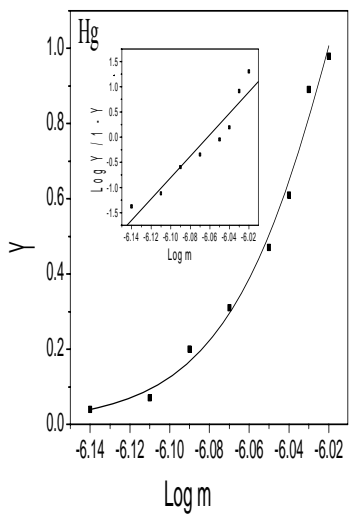

**Fig.4.** Binding Isotherm of Hg(II) ions binding with C-PC in the Scatchard coordinates (dependence of *y* against *log m*, where *y=r/n* and *r=c$_{bound}$*/[C-PC] is concentration of bound metal ions per mole C-PC, *m* is concentration of free metal ions, *n* is  r$_{max}$) (R 0.99). insert:  Hill plot  (R 0.98) (parameters are the same, as in the case Scatchard analysis).

The binding constants estimated from these isotherms are shown in Table 3.

**Table 3.** Comparative energetics of metal ions binding with C-PC at 23°C

| | | Metal ion | Hg(II) | Ag(I) | Pb(II) | Cr(III) | Cu(II) |
|---|---|---|---|---|---|---|---|
| Fluorescence titration | Scatchard analysis | Constant of binding $K \times 10^5 M^{-1}$ | 11.4±1.0 | 5.2±0.5 | 2.8±0.3 | 1.0±0.1 | 0.6±0.1 |
| | | Gibbs free energy $-\Delta G°$, kcal/mol | 8.2 | 7.8 | 7.4 | 6.8 | 6.5 |
| | Hill analysis | Constant of binding $K \times 10^5 M^{-1}$ | 12.0±1.3 | 5.5±0.9 | 2.3±0.4 | 0.90±0.08 | 0.74±0.15 |
| | | Gibbs free energy $-\Delta G°$, kcal/mol | 8.3 | 7.8 | 7.3 | 6.8 | 6.6 |

The results of the study indicate that value of calculated $\Delta G°$ of Cu(II) and Cr(III) ions is characteristic for hydrogen bonds (6.8÷6.5 kcal/mol) whereas Ag(I) and Pb(II) are almost on the verge of (7.4÷7.8 kcal/mol) and Hg(II) exceeds the energy of hydrogen bonding (8.2 kcal/mol) (table 3).

Thus, binding affinities of toxic and heavy metal ions for C-PC strongly depend upon the metal and the constants of binding are arranged in the descending order as follows:

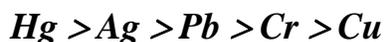

$$Hg > Ag > Pb > Cr > Cu$$

As it can be seen from the fluorescence data, the values of the binding constant $K$, determined both by the Scatchard and the Hill graphical analysis, are in good agreement. The first question which was asked what kind of site on a protein will a given metal ion select for binding.

C-phycocyanin is high organized globular protein (hexamer). Monomer of C-PC consists from α- and β- polypeptide chains with 3 covalently binding chromophores (one is on the α- chain and two are on the β-chain) Whereas tetrapyrrol group (phycocyanobilin) in C-PC is chromophore, the possibility of complex formation between metal ions and chromophore of C-PC can be supposed. It is known from the literature data that metal preferentially bind with thiol groups of proteins. Naturally, the metal effect depends also on the structural position of the chromophore in the protein. In phycocyanobiline the main prosthetic group of C-PC is fixed in α-84 and β-84 and β-155 positions by cystein thioester groups[9,10]. It was shown by applying the Forster theory [11] that β-84 chromophore is fluorescent[12]. Proceeding from our data on C-PC fluorescence quenching by metal ions, it is quite possible that cystein in β-84 position is one of the specific binding sites for metal ions in C-PC. At the same time, hexamers of α and β-subunits form globule mostly, so in C-PC the negative electrostatic field of hexamer is dominant. It is quite possible that such a distribution of the charge also plays an important role in the interaction with metal ions.

Thus, there are two possible energetically distinct modes of metal ions binding to C-PC: site-specific and nonspecific (diffuse) binding. Diffusely bound toxic metal ions are electrostatically attracted to the strong anionic field in the center of the hexamer. Site-specifically bound metal ions are strongly attracted to specifically arranged ligands-thiol groups of C-PC.

If electrostatic interactions occur between the protein and a charged quencher, an additional approach to gain insight concerning the local electric potencial in the vicinity of the fluorophore is by studying the ionic strength dependence of the binding efficiency.

In this section the energetics of Cu(II) ions interaction with C-phycocyanin from *S. platensis* is defined via determination of the stoichiometric constants by equilibrium dialysis at different ionic strengths (fig.5).

To date, very few methods allow direct determination of protein-metal interactions, as well as exact stoichiometric binding ratios. For determination the number of active binding sites (n) were used equilibrium dialysis experiments. During equilibrium dialysis on the one side of the semipermeable membrane is C-phycocyanin, to which the membrane is impermeable, and on the other side are metal ions, which diffuse through the membrane. After attaining equilibrium (in this case, the dialysis lasted 72 h), a part of metal ions remain free and the other part are bound to C-PC, with the principal balance naturally maintained: *[Metal]$_{total}$ =[Metal]$_{free}$ + [Metal-C_PC]*. It proved that this balance is disturbed at high metal concentrations:

$$[Metal]_{total} > [Metal]_{free} + [Metal\text{-}C\_PC]$$

Even at concentrations of [*Metal*]/*C_PC>10, there was* noticeable salting out on the membrane. Therefore, we defined *n* as the limiting value of *r* at which the principal balance is maintained.

As is seen from the table 4, the number of binding sites varies in the interval 3÷4 i.e. about 4 mol of Cu(II) falls per mol of C-PC. Cu(II)–C-PC (using equilibrium dialysis) binding isotherms are given in the Scatchard coordinates at different ionic strengths in fig.5. The points represent experimental data whereas solid curves are hypothetic functions fitted by $\chi^2$ criterion. Binding constants were estimated by these isotherms (table 4). Comparison of *K* values at different ionic strengths shows that within the error limits they do not differ virtually at 0.02M and 0.05M, whereas at 0.002 M NaCl $K_{0.002\ M\ NaCl}$ is greater than $K_{0.02\ M\ NaCl}$ and $K_{0.05\ M\ NaCl}$ i.e. *K* decreases with ionic strength increase.

The type of *y* versus *logm* dependence means that there exists positive cooperation of interaction of metal ions bound with C-PC, i.e. binding of the first metal ion increases affinity of the site for the second one. To illustrate the possibility, the obtained data were additionally plotted in the Hill coordinates. The following binding parameters were determined: the binding constant *K*, the Hill coefficient $n_H$, which is the cooperativity index. All the results are given in the table 4. As it is seen, the binding constants *K* determined by both the Scatchard and Hill methods are in good agreement, which is also valid for *ΔG°*. *ΔG$^0$* is approximately 7 kcal/mol that is characteristic for hydrogen bonds. The dependence of Cu(II)-C-PC complex stability on ionic strength of solution shows the competition of sodium Na(I) and Cu(II) ions for binding sites.

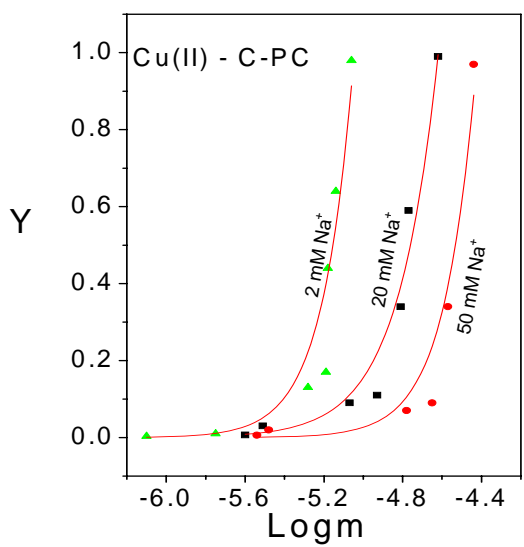

**Fig. 5**. The binding isotherm of Cu(II) ions with C-PC at various ionic strengths. (The dependence *Y vs Log m*, where $Y = r/n$ and $r = C_{bound}/$ [C-PC] is the concentration of bound metal ions per mol C-PC, m is the concentration of free metal ions, n is the number of binding sites for the metal ions per mol C-PC at saturation) ($\chi^2$ 0.005 ÷ 0.03). Each point represents the mean from three independent determinations and standard deviations are < 12 % of the means.

**Table 4.** Parameters of Cu(II) ions binding with C-PC, at 20°C.

|  | Ionic strength | 0.05 M | 0.02 M | 0.002 M |
|---|---|---|---|---|
| Analysis by the Scatchard method | Binding constant, $K \times 10^4$ M$^{-1}$ | 7.41 | 8.1 | 10.98 |
| | Gibbs free energy, $-\Delta G^0$ kcal/mol | 6.62 | 6.68 | 6.86 |
| | Number of binding sites *n* | 4.1 | 3.94 | 3.54 |
| Analysis by the Hill method | Binding constant, $K \times 10^4$ M$^{-1}$ | 7.2 | 8.2 | 11.2 |
| | Gibbs free energy, $-\Delta G^0$ kcal/mol | 6.6 | 6.68 | 6.87 |
| | Hill coefficient $n_H$ | 3.1 | 2.3 | 2.57 |

The dependence of $pK$ (for Me-C-PC complexes using absorption titration) upon the covalent index $X_m^2 r$ is presented in Fig.6. Correlation is observed between $pK$ and the covalent index $X_m^2 r$. ($X_m$ Pauling electronegativity, r ionic radius corresponding to the most common coordination number). $X_m^2 r$ is a quotient that compares valence orbital energy with ionic energy. It is a measure of the relative ability of metal ions to participate in covalent interactions compared to ionic interactions[13]. The correlation between theoretical covalent index and empirical binding constants clearly demonstrate the dynamic nature of metal-PC interactions.

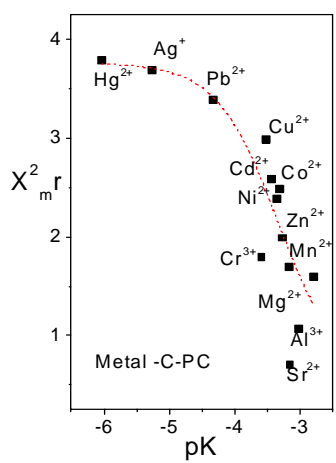

Fig.6. Correlation between the covalent index $X_m^2 r$ and binding constant logarithm for metal-C-PC complexes (R 0.9).